\documentclass[a4paper]{IROS}
\usepackage[latin1]{inputenc}
\usepackage{amsmath}
\usepackage{amssymb}
\usepackage[dvips]{graphicx}

\begin{document} 

% create the title header:

\title{Multi-Modal Human-Machine Communication for Instructing Robot Grasping 
Tasks\footnote{{\it appeared in}: Proceedings of the IEEE/RSJ International Conference on Intelligent Robots and Systems (IROS), Lausanne, Switzerland, IEEE publications, pp. 1082-1089 (2002)}}
\author{P. McGuire, J. Fritsch, J. J.  Steil, F. Röthling, G. A. Fink, S. Wachsmuth, G. Sagerer, H. Ritter}
\affiliation{Bielefeld University, Faculty of Technology\\
  P.O.-Box 10 01 31, D-33501 Bielefeld, Germany\\
 \{mcguire,jannik,jsteil\}\symbol{64}techfak.uni-bielefeld.de} \date{}
\maketitle\thispagestyle{empty} % don't forget \thispagestyle{empty}, otherwise you'll get page numbering

% write the abstract with the Abstract-environment:

\begin{Abstract} A major challenge for the realization of
  intelligent robots is to supply them with cognitive abilities
in order to allow ordinary users to program them  easily and intuitively. 
One way of such programming is teaching work tasks by interactive demonstration. 
To make this effective and convenient for the user, the machine must be capable
 to establish a common focus of attention and  be able to use and integrate spoken
 instructions, visual perceptions, and non-verbal clues like gestural commands. 
We report progress in building a hybrid architecture that  combines statistical
 methods, neural networks, and finite state machines into an integrated system for
instructing grasping tasks  by  man-machine interaction. The system  combines 
the GRAVIS-robot for visual attention and gestural instruction with  an intelligent 
interface for  speech recognition and linguistic interpretation,  and 
an modality fusion  module to allow multi-modal task-oriented man-machine 
communication with respect to dextrous robot manipulation of objects. 

\end{Abstract}

% ...and start writing!

\section{Introduction}  
\begin{figure}[b!]
\begin{center}
 \includegraphics[width=\columnwidth]{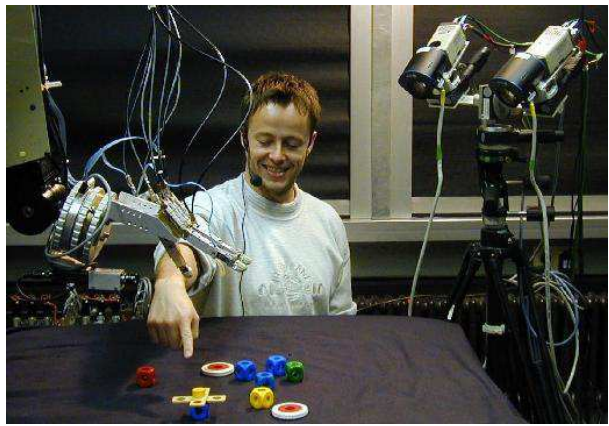}
\caption{The interactive scenario.}\label{Fig:Scene}
\end{center}
\end{figure}

In recent years a  new generation of intelligent robots   
 has found applications in natural environments like museums, hospitals, or private
households. While conventional programming can be efficient for factory floor
applications, more cognitively oriented robots  must be instructable by ordinary
human users in a robust and intuitive  way. In this respect, one 
 way to program a work task is  by interactive human demonstration, 
which requires the endowment of  a robot with  sufficient perceptual,
 cognitive, and motor skills  to communicate with the
 user in a natural fashion. As humans inevitably use different modalities
 in interpersonal and man-machine communication, an intelligent robot system
 should take advantage of this information by using and  integrating 
 different perceptual channels. In this paper, we present  a combination and
 integration  of active vision, gestural  instruction, and speech input to 
instruct a robot system for grasping tasks (Fig.~\ref{Fig:Scene}). Though parts of the functional modules
have  been described and evaluated as standalone applications in more detail
 earlier \cite{Bauckhage2001-AIS,BrandtPook99:IRA,JungclausRaeRitter1999-AIS,SteilHeidemannJockuschRaeJungclausRitter2001-GAF}, their integration  into a  full scale architecture  is described here 
for the first time and
has proven to be a major challenge due to the enormous complexity of the overall system.
Therefore we focus on the architecture and  module interconnections
 and highlight some lessons learnt from building such an  interactive system. 
As a  whole, the described project is part of a larger research
effort (Bielefeld Special Collaborative Research Unit SFB 360 \cite{RickWachs})
aiming towards the development of ``situated artificial
communicators'' that can be interacted with in a natural, ``human-like''
fashion with the combined use of  verbal and non-verbal  instructions.
It is in line with earlier work devoted to robot teaching by showing
 \cite{Kuniyoshi94} and imitation learning \cite{Bakker96,Breazeal00}.
While there has been much work on various aspects of learning in cognitive 
architectures  (speech and image integration \cite{McKevitt94:ION}, 
trajectory acquisition \cite{Dillmann95,IjspeertSchaal2001,Yeasin97},
object recognition and grasp pose determination \cite{Pauli98} or sensor 
fusion for grasping \cite{Allen1999}), the design of 
an integrated architecture is  widely believed to
be very   hard to achieve. Thus there have been developed only a  
few advanced architectures which are 
capable of integrating perceptual attention mechanism with higher level
functions \cite{BreazealScassellati1999,BrooksXX,Ghidary2001}.

The next sections provide an overview of the overall system and its
highest level  building blocks, with special emphasis on their mutual interactions. 
We then demonstrate some of the system's capabilities and discuss and illustrate
the idea that there exists  a ``critical level of skills'' from which 
development of the system towards  more complex capabilities progresses much
faster. 

\section{System Architecture}
The architecture design is one of the key issues in realizing a complex
intelligent robot system.  From an ideal perspective, a common uniform 
software framework should be specified beforehand to support a subsequent
distributed development of modules according to certain specifications. 
Different approaches like behavior based architectures, agent-based concepts 
or blackboard systems have been proposed in this context. 

 However, in a truly complex system very different types of signals are
 generated at different time  scales and require  many sub-skills 
 to be developed under diverse programming paradigms.  In Section 8 we discuss 
further reasons why from our experience  it is unreasonable to 
impose strong constraints  on the submodules for easier software engineering. 
As a consequence, we find that it  is rather the level of the architecture  
which has to support  the integration of heterogeneous components. 

Our  entire system is implemented as a larger number of separate
processes running in parallel on several workstations  and
communicating with the distributed architecture communication system (DACS
\cite{Fink95:DACS}) developed earlier for the purpose of this project. 
Hereby the submodules use different programming languages (C,C++,Tcl/Tk,Neo/NST), 
various visualization tools, and a variety of processing paradigms ranging
from a neurally inspired attention system to statistical and declarative methods
for inference and knowledge representation. Thus the architecture as 
a whole cannot be easily subsumed under any single one  of the programming  paradigms
 mentioned above.

\begin{figure}[t!]
\begin{center}
 \includegraphics[width=\columnwidth]{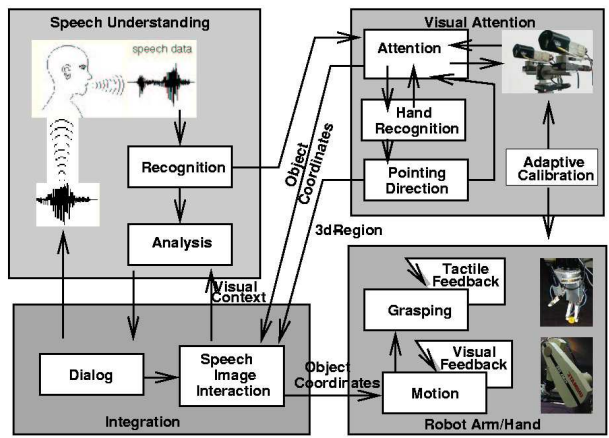}
\caption{Modules of the integrated architecture.}\label{Fig:System}
\end{center}
\end{figure}

Figure~\ref{Fig:System} shows a coarse  overview  of the main 
information processing paths. The speech processing (left) and the attention
mechanism (right) provide linguistic and visual/gestural inputs converging 
in an integration module which then passes control to  the manipulator. Additionally, 
there are  control commands for  parts of the system
(e.g. on, off, calibrate skin, park robot arm,...). The modules and some
of their interactions  are further described in the following sections.

\subsection{Hardware Basis }
The vision hardware
 currently consists of a binocular active vision head with two
3-chip-CCD color-cameras, controllable pan, tilt, left/right vergence and
motorized lenses determining focus, zoom and aperture,   which combine to 
 a total of 10 DOFs. The
grasping and manipulation is carried out by a standard 6DOF PUMA manipulator
operated with  the real-time RCCL-command  library. It is additionally equipped with 
a wrist camera to obtain local  visual feedback during the grasping phase. 

Grasping is carried out by a 9DOF dextrous robot hand 
developed at the Technical University of Munich. It has three
approximately human-sized fingers driven by an oil hydraulics system. 
The fingertips have  custom built fingertip sensors to provide force feedback for 
control and evaluation of the grasp. The hardware setting  and its 
 control design has been described in more
detail  in \cite{SteilHeidemannJockuschRaeJungclausRitter2001-GAF}. Recently 
we have changed the original hand design  by 
adding a palm and rearranging  the fingers in a more human-like configuration 
(Fig.~\ref{Fig:Scene},\ref{Fig:Grasp})
 to allow a larger variety of two- and three-finger grasps.

\section{Visual Attention and Memory}
A necessary prerequisite for successful human-machine interaction is to
establish and maintain a common focus of attention between the user and 
the vision system of the robot. Furthermore,  a  short term visual memory 
has to be realized in order  to understand linguistic  reference to 
objects in spoken instructions.  Our attention system  places a high emphasis on the 
spatial organization of  visual clues and enhances a design proposed  in
\cite{JungclausRaeRitter1999-AIS}. It  consists of 
a layered system of topographically organized neural maps
for integrating different low-level feature maps into a continually
updated focus of attention for the active camera head. Similar mechanisms have 
also been employed   in
\cite{BreazealScassellati1999,Driscoll98,Sethu2001},
 however, only results for highly idealized synthetic 
 images or  using  a lower number and  less complex maps are reported.

\begin{figure}[t!]
\begin{center}
 \includegraphics[width=\columnwidth]{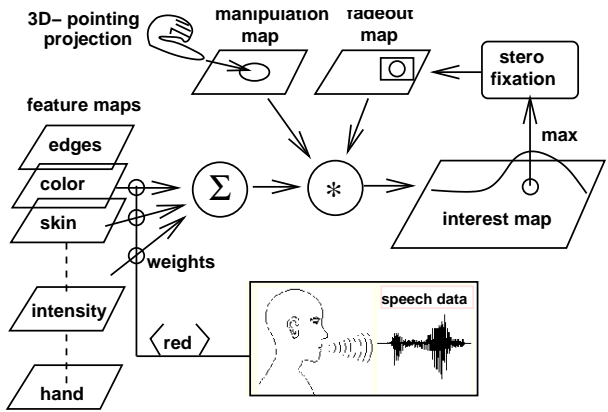}
\caption{User speech input  (``take the red ... '') can bias the attention
  system towards special features (red) and  3D-pointing gestures impose
  constraints  for spatial interest regions.}\label{Fig:Attention}
\end{center}
\end{figure}

In particular, from the stereo images 
a number of feature maps indicating the presence of 
oriented edges, HSI-color  saturation \& intensity, motion (difference map), 
and skin color are computed. As one of the main goals of the system is to
recognize pointing hands, we multiply the difference map (indicating movement) by the
skin segmentation map (indicating a hand). The result is a ``moving skin''
map, which is considered as  a separate feature map.  
A weighted sum of these feature maps is multiplied by  a fadeout-map to form a
final attention map  and the highest peak determines the 
next fixation, see Fig.~\ref{Fig:Attention}. After stereo matching, 
the resulting loop  continuously generates saccades for  fixations and this
 active exploration behavior persists during the whole system operation. 

%\subsection{Interactive attention biasing}
Interaction with the human user can modify the attention map  by two 
different mechanisms. If a spoken instruction references a colored object (``
... the red cube ...'') the corresponding weight is increased to bias the
attention system towards red  spots in the image.
This  increases the probability for fixations on red things, but after some 
time a  decay mechanism drives the weighting back to a default
level. 

If the  hand and gesture recognition modules
 detect a pointing gesture in the image,  the 3D-direction of the pointing 
finger is  computed and a corresponding region of interest is virtually
 projected on the table. A respective ``manipulation map'' is 
 multiplied coordinate-wise with the attention map 
to restrict the explorative attention  to that region in the next step, see
 Fig.~\ref{Fig:Attention}. 
 
The exploration behavior tends to fixate repetitively upon the most
interesting  points, which are in most cases objects. This ``emerging regularity''
is used  to establish a short term visual 
memory in the integration module to which all 3D-fixation  coordinates are sent. 
It  uses  temporal integration to stabilize only the 
most salient points and if additionally a homogeneous color blob is detected, 
it is assumed that there is an object, which then can be referenced
by spoken instructions. Future extensions will add a more sophisticated object
recognition (already available for the grasping feedback) at this point. Also 
we plan to add more specific object maps in the attention system, which then
can be favored by spoken instructions exactly like the color maps. 

%\section{Speech and Language}

 %\input{speech}

\section{Speech and Language}

%A vision system does not stand for its own. 
%In our scenario it enables
%a human instructor to reference on objects in the scene in a natural way.
%In our scenario a human may instruct the system to take some parts, connect them together,
%add another part, and put the assembly back into the scene. 
%
%In our scenario verbal object descriptions have to be related with observable
%objects. 
%In order to realize this task in a {\em natural way},
%the user should not adapt to the system, but the system has to be designed
%for the user. 
To allow a fluent communication between the instructor and the
artificial communicator 
our system is capable of understanding speaker independent speech input.
The instructor neither needs to know a special
command syntax nor the exact terms or identifiers of the objects.
Consequently, the complete speech understanding system has to face a high degree of
{\em referential uncertainty} from vague meanings, speech recognition errors,
and un-modeled language structures.
%, and, on the visual input channel,
%erroneous object classification results.

%In the context of a multi-modal scene understanding system 
%speech is interpreted as a second source of information describing the 
%visually perceived scene.
%Therefore, special attention is given to verbal object descriptions.
Our approach to robust spoken language understanding uses
a vertical organization of
knowledge representation and an integrated processing scheme to overcome
the drawbacks of the traditional horizontal architecture \cite{BrandtPook99:IRA}.
%The resulting tight coupling of speech recognition and
%understanding is shown in Fig.~\ref{speech_overview}.
%\begin{figure}[htb]
%  \includegraphics[width=\columnwidth]{speech_system.eps}
%  \caption{Overview of the integrated speech interpretation.}
%  \label{speech_overview}
%\end{figure}
As baseline module \cite{Fink99:DHM} it employs an enhanced  statistical speech
recognizer.
The recognition process is directly influenced by a partial parser
which provides linguistic and domain-specific restrictions on word sequences.
Therefore, partial syntactic structures  
instead of simple word sequences are generated, like
e.g. object descriptions
("the red cube") or spatial relations ("...in front of...").
These are combined by the subsequent speech understanding module 
to form linguistic interpretations. 
%Although there has been some progress recently,
%the detection of {\em out-of-vocabulary} words can still not be performed
%robustly on the level of acoustic recognition. Therefore, 

To cope with {\em out-of-vocabulary} words
we employ a
recognition lexicon which exceeds the one used by the understanding component
but covers all lexical items frequently found in our corpus of human-human
and human-machine dialogs. 
The syntactic modeling then allows one to use these additional words to
be filled-in for such open lexical categories as nouns, for example.
In a robust system the speech processing modules have to be able to 
cope with spontaneous speech input which largely deviates from speech 
read from  text prompts or used in a dictation task. 
Particularly,  clear pronunciation, vocabulary limitations,
and restrictions in language-use can never be enforced.
%Since spontaneous speech input largely deviates from speech 
%read from e.g. text prompts
To meet these challenges the recognition
lexicon contains acoustic models for spontaneous speech phenomena, namely for 
so-called \textit{human noises} (breathing or lip smacks) and hesitations
(like '\texttt{uhm}').

%\section{Integration}
%\input{integration}

\section{Integration}

\subsection{Interrelating Speech and Vision}

If a naive user describes an object in the scene by using attributes he or she will
typically use a vocabulary which is different from the fixed one  appropriate
for processing of visual data. Therefore, several kinds of uncertainties 
have to be considered when correlating a verbal object description and
object recognition results, such as vague attributes (e.g. {\em ``the long,
thin stick''}), vague spatial and structural descriptions 
(e.g. {\em ``the object to the left of the cube'', ``the cube with the bolt''}),
or speech and object recognition errors.

\begin{figure}[h!]
\begin{center}
 \includegraphics[width=\columnwidth]{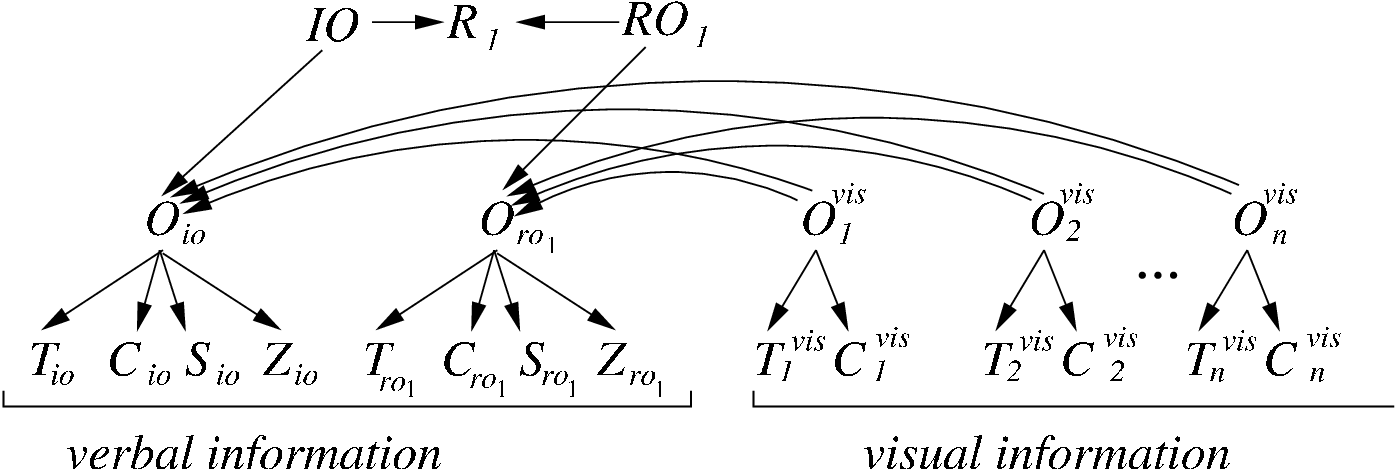}
\caption{Bayesian network connecting 
  speech and vision data for one related reference object.}
\label{speimg}
\end{center}
\end{figure}
In order to cope with this uncertainty, we have developed a Bayesian network
approach that robustly combines verbal and visual information through
different abstraction levels \cite{Wachsmuth99:MIO}. 
On the first level, basic features
from vision ({\em $C^{vis}$ :color, $T^{vis}$ :elemental type}) 
and speech ({\em $T$ :type, $C$ :color, $S$ :shape, $Z$ :size}) 
are modeled as evidential nodes of the Bayesian network for each of
the $n$ visual objects and $N$ verbally-referenced objects.
On the second level, these are fused to the {\em visual object class} 
$O^{vis}_{k\in\{1,\dots,n\}}$ and 
{\em verbal object class} $O_{io/ro_j,j\in \{1,\dots,N-1\}}$ which are
connected by the intended object and reference object variables 
$IO,RO_j\in\{1\dots n\}$. The verbally-mentioned spatial or structural 
relations between the objects are established
by introducing additional evidential nodes $R_j$ (Fig.~\ref{speimg}). 
The different kinds of uncertainties are modeled by conditional probability
tables that have been estimated from experimental data \cite{Wachsmuth99:MIO}.
The objects which are denoted in the utterance 
are those explaining the
observed visual and verbal evidences $e^{vis}, e^{verb}$
in the Bayesian network with the maximum a posteriori
probability. Additional causal support for an intended object IO is defined by
an optional target region of interest that is provided from the 3D-pointing
evaluation. The intended object $IO$ is then used by the dialog component for system
response and manipulator instruction.

\subsection{Dialog system}

%In the last years, many spoken dialog systems have been developed
%(e.g. \cite{Os99:OOT,Rudnicky99:CND}).
%However, most of these systems 
Many dialog systems developed recently 
lack  integration with other modalities.
In contrast to 
such uni-modal
%these 
approaches 
our dialog module integrates utterances
of the instructor, information of the visible scene, and feedback from
the robot to realize a natural, flexible and
robust dialog strategy.

\begin{figure}[t!]
\begin{center}
 \includegraphics[width=\columnwidth]{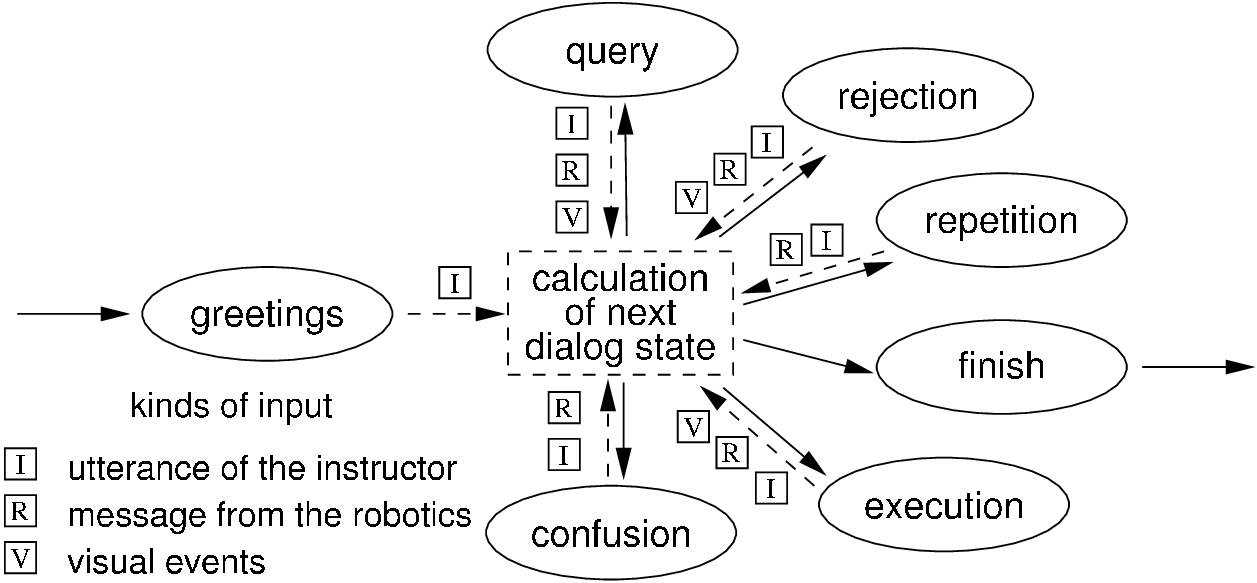}
  \caption{Model of the man-machine communication.}
  \label{dia_ablauf}
\end{center}
\end{figure}

The dialog module is realized within the semantic
network language \textit{Ernest} using the dialog
model shown in Fig.~\ref{dia_ablauf}. 
The model is based on an investigation of a corpus of human-human and simulated
human-machine dialogs. 
Every path through the model reflects a course of a possible human-machine
dialog. 
%The dialog starts with a greeting generated by the module
%and ends, if the instructor indicates the end of the assembly process.
The admissible sequence of intermediate states is nearly
unrestricted leading to a very natural and robust dialog behavior 
\cite{BrandtPook99:IRA}.

State transitions are initiated if new information from the
instructor or
%, if connected to the system, 
the robot is available.
The state transition function analyzes the new information and combines it with the current
dialog context and information gathered from the interrelation module to
select the next state.

Using the dialog context, references between objects can be resolved,
 {\em ``Take the red bolt. Put \emph{it} into the cube.''},
and information accumulated in the dialog can be combined.
The dialog module can react upon new information from different
modalities to  inform the instructor about errors
during the execution of an action and can actively control the dialog to
 query for missing or unprecise information.
%In order to actually execute an action references
%between verbally described objects and objects in the scene must be
%established. This references are obtained from the integration module,
%which yields all possible references between one verbally described
%object and all results from the vision component. All references,
%whose verbal and visual descriptions support each other, are used in
%the further processing.
The overall goal of this module is to continue the dialog in every
situation. Actions which cannot be executed are immediately
rejected. For verbal instructions which could not be analyzed a
repetition is requested up to two times. If the dialog has gathered
completely contradictory information the system expresses its confusion and
asks for a new instruction.

\begin{figure}[t!]
\begin{center}
 \includegraphics[width=\columnwidth]{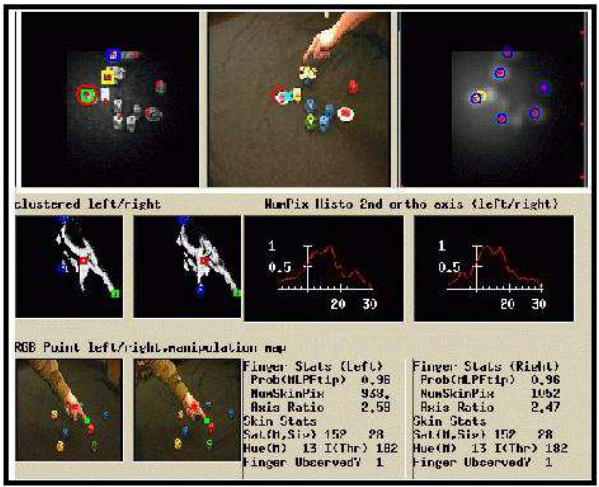}
\caption{The attention map:  hot spots, camera image, stereo matched points to
  be transferred to the integration module (upper row). Hand and finger
 recognition uses a multi-layer perceptron based classification of the intensity
 histograms (middle row). Projection of the 3D-pointing direction on the table
 (lower row).}\label{Fig:Gest}
\end{center}
\end{figure}

\section{Manipulation}

Once the integration module has resolved ambiguities, 
 control is passed to the robot arm/hand. 
Starting from the 3D-coordinates  determined by the vision and integration modules 
 the approaching movement and  grasping 
is  executed in a semi-autonomous fashion relying on 
local feedback only. The arm and hand control is implemented
as a  finite state automaton, switching between different arm modes
(approach, refine, closer, re-align,...) and hand states (open, pre-shape, grasp, hold,
release,...)  whose transitions are triggered by visual and tactile feedback. 
In particular, the wrist camera provides visual feedback  
and object recognition to approach the grasp offset position and, 
in  the  grasping phase,  the fingertip sensors provide the necessary force feedback. 

The grasping  sequence starts with an approach movement, 
recenters the manipulator above the object,
 chooses a grasp prototype according to the recognized object, aligns
the hand along the main axis of the object and executes the grasp prototype, 
for more details see \cite{SteilHeidemannJockuschRaeJungclausRitter2001-GAF}.
After successful gripping, a similar chain of events allows the robot to put the
object down in another gesturally selected location.

\section{An action sequence }
\begin{figure}[b!]
\begin{center}
 \includegraphics[width=\columnwidth]{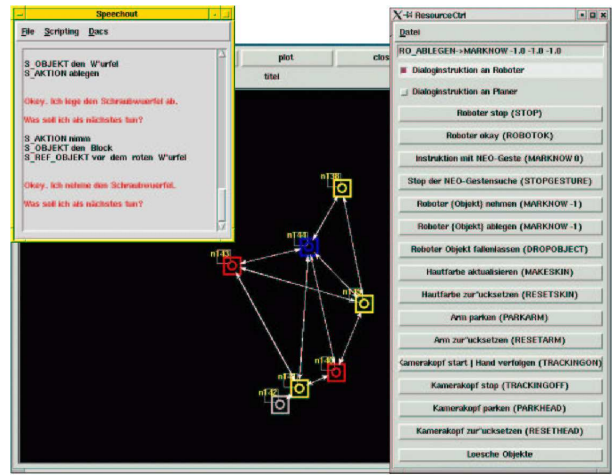}
\caption{The speech input  is analyzed and  segmented into semantic
categories like action (S\_AKTION nimm) or object (S\_OBJEKT den Block)
(upper left display). Spatial relations between objects in the  short time 
memory (lower display) can be referenced by instructions. In  the right window
a number of direct commands are available, which can also be given by spoken 
instructions. }\label{Fig:Objmem}
\end{center}
\end{figure}

\begin{figure}[t!]
\begin{center}
 \includegraphics[width=\columnwidth]{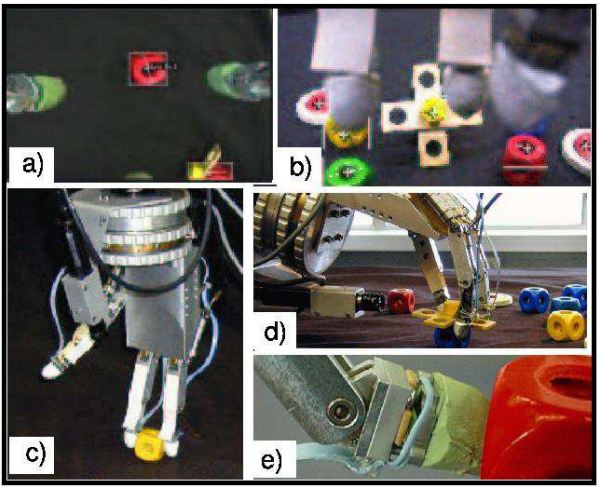}
\caption{Visual and tactile feedback for grasping: a) view through the 
hand camera for the approach movements; b) two finger grasp of a cube; c)
hand camera view before three finger grasp;d) three finger grasp; e) force
feedback from the fingertips to evaluate the grasp.}\label{Fig:Grasp}
\end{center}
\end{figure}

To illustrate the capabilities of our system  we present a (simplified) 
typical action sequence  for picking up and deploying  an object in sequential order. 
Some videos can be found at \cite{GRAVIS-home}. The sequence consists of 8 major stages:\\
1) Initially, a number of objects are spread on a table in the workspace of
robot arm  and camera. The system can be started and partially calibrated 
by speech commands shown in  Fig.~\ref{Fig:Objmem} (right display). 
 The attention systems explores the scene as shown in Fig.~\ref{Fig:Gest} (top row)  and transmits
the fixation points to the integration module, where the visual memory is 
stabilized and the spatial object relations  are analyzed, see
 Fig.~\ref{Fig:Objmem} (lower left display).\\ 
 2) A user gives a spoken instruction referencing
one of the objects. The instruction is semantically analyzed and the 
dialog is initiated, see Fig.~\ref{Fig:Objmem} (upper left display). The system may ask for 
additional pointing information, e.g. for resolving ambiguities. It also 
determines, whether the attention system should be biased towards particular colors. \\
3)  When a pointing hand is found, the gesture is 
evaluated as visualized in Fig.~\ref{Fig:Gest} (middle row) and 
 the 3D interest region is fed  to the integration module.  \\
4) The Bayesian network integrates the spoken instruction, the visual memory, and 
the gesture-based region bias to determine the object to be grasped. In case
this fails, the dialog asks for a repetition of the instruction and a new gesture.\\
5) Control is passed to the hand/arm system, which performs a visually guided 
approach movement (Fig.~\ref{Fig:Grasp}(a)), determines a grasp primitive and 
pre-shapes the hand (c), aligns it
with respect to the object and finally grasps the object (b,d))  with  force feedback
control (e).  Upon a failure, it retries and on success the integration module
is informed. \\
6) The dialog system  asks the user to indicate where to deploy the object. \\
7) The pointing evaluation part of 4) is repeated with the slight difference
that now the 3D-fingertip position directly determines the position of object deployment. \\
8) Control is redirected to the hand/arm system, which deploys the object and 
the system returns into  the starting mode of exploration.

\section{The Critical Level of Skills}

As described above, our system integrates a larger number of skills, local 
feedback mechanisms, and state machines. 
Many of the modules have been developed and tested
independently of each other, can be  trained offline, 
and have adaptive calibration facilities \cite{Bauckhage2001-AIS,SteilHeidemannJockuschRaeJungclausRitter2001-GAF,Wachsmuth99:MIO}. Most of  them
are much more powerful when operated standalone; however,  in the integrated  
system, the full capabilities of each module are not always employed. This is 
 due to mutual interdependencies and  less specialized hardware delivering 
a lower quality of sensory inputs. Consequently, a potential of
``hidden capabilities'' and resources exists in the overall system. 

One  approach to avoid  this apparent waste of capabilities is to 
restrict the solution space for the individual modules 
to ensure a high degree of homogeneity  towards a  beforehand specified
scenario. However, we find that this  is not reasonable 
because  once a robust functioning of the overall system is achieved,
we  can  benefit  from the ``hidden capabilities'' in certain modules
 quite easily. We experience that small
 coordinated modifications in several modules or slight changes in 
the control flow can quickly  open new and unforeseen perspectives for 
the system. We give two examples to  illustrate this: Recognition of bars
together with a corresponding grasp prototype allows us
to progress from cube-based pyramid building to cube and 
bar based building of  bridges, houses, closed boxes, etc.. Secondly, a slight change in the
speech-initiated control 
allows  to reuse the  fingertip detection algorithm
initially employed  to find  objects  for deploying them
 at fingertip positions. The same capabilities can be used to teach multi-point
trajectories just by pointing to  consecutive positions  or to indicate small
relative movements by pointing to two nearby positions subsequently. 

We believe that this  experience can be summarized as approaching  a
 \textit{critical level of skills}. This level  is 
characterized by a situation where 
small improvements or (adaptive) reconfiguration of single modules 
or slight changes in the control flow
immediately open up a whole new variety of action  opportunities. 
Hereby we benefit from a certain amount of robustness, the possibility
to readapt or recalibrate,  and a rather loose coupling between the modules, which 
in our architecture is realized by the message-passing communication 
paradigm and which allows quick reorganization of the control flows. 
 The interactive teaching of tasks then can 
take full advantage of the user's creativity to recombine the system's skills
towards previously unexpected results. 

\section{Discussion} 

The presented architecture integrates a set of capabilities to enable an intuitive 
programming of grasping  tasks by a human user. It ranges from 
a perceptual grounding in an active exploration of the scene up to an 
interpretation  of complex user commands by a sophisticated speech analysis 
and modality fusion system. As there are no widely accepted 
benchmarks for cognitive robotic systems interacting with humans, 
it is  difficult to assess  the performance of such  systems systematically and 
beyond  demonstrating  that they  are indeed running  by examples. 
Thus, currently we are adding  a  visualization and monitoring module, which 
will also be able to record action sequences and will  enable a more   
quantitative performance analysis.

We think one of the major challenges is to lift learning  in our system 
from the  offline training widely used in  the lower level  modules to the level 
of behavior. The current system,  enhanced by a system 
monitor,  will offer a  tool to study how such learning needs to be organized 
to progress from imitation of human-instructed action sequences to extracting 
knowledge on the task level. Some  of the  many issues will be how to propagate errors
top down and how to flexibly reorganize the control flow without losing
robustness and functionality of the system. Only then will we come closer to 
easily-instructable intelligent systems that can robustly carry out non-trivial 
tasks in natural environments.

% and the bibliography:

%\begin{thebibliography}{99}

%\end{thebibliography}

\bibliographystyle{plain}

\end{document}